\documentclass[aip,jap,reprint,superscriptaddress]{revtex4-1}

\usepackage{graphicx}
\usepackage{amsmath}
\usepackage[colorlinks=true,citecolor=blue,linkcolor=blue]{hyperref}
\hypersetup{colorlinks = true, linkcolor = blue, urlcolor=blue, bookmarksnumbered =  true}

\begin{document}
%Title of paper
\title{Photonic nano-structures on $\left(111\right)$-oriented diamond}
\author{Elke Neu}
\author{Patrick Appel}
\author{Marc Ganzhorn}
\author{Javier Miguel-S\'{a}nchez}
\affiliation{Department of Physics, University of Basel, Klingelbergstrasse 82, CH-4056 Basel, Switzerland}
\author{Margarita Lesik}
\affiliation{Laboratoire Aim\'{e} Cotton, CNRS, Universit\'{e} Paris-Sud and Ecole Normale Sup\'{e}rieure de Cachan, F-91405 Orsay, France}
\author{Vianney Mille}
\affiliation{Universit\'{e} Paris 13, Sorbonne Paris Cit\'{e}, Laboratoire des Sciences des Proc\'{e}d\'{e}s et des Mat\'{e}riaux (CNRS UPR 3407), F-93430 Villetaneuse, France}
\author{Vincent Jacques}
\affiliation{Laboratoire Aim\'{e} Cotton, CNRS, Universit\'{e} Paris-Sud and Ecole Normale Sup\'{e}rieure de Cachan, F-91405 Orsay, France}
\author{Alexandre Tallaire}
\affiliation{Universit\'{e} Paris 13, Sorbonne Paris Cit\'{e}, Laboratoire des Sciences des Proc\'{e}d\'{e}s et des Mat\'{e}riaux (CNRS UPR 3407), F-93430 Villetaneuse, France}
\author{Jocelyn Achard}
\affiliation{Universit\'{e} Paris 13, Sorbonne Paris Cit\'{e}, Laboratoire des Sciences des Proc\'{e}d\'{e}s et des Mat\'{e}riaux (CNRS UPR 3407), F-93430 Villetaneuse, France}
\author{Patrick Maletinsky}
\email[]{patrick.maletinsky@unibas.ch}
\affiliation{Department of Physics, University of Basel, Klingelbergstrasse 82, CH-4056 Basel, Switzerland}

\date{\today}

\begin{abstract}
We demonstrate the fabrication of single-crystalline diamond nanopillars on a (111)-oriented chemical vapor deposited diamond substrate. This crystal orientation offers optimal coupling of nitrogen-vacancy (NV) center emission to the nanopillar mode and is thus advantageous over previous approaches. We characterize single native NV centers in these nanopillars and find one of the highest reported saturated fluorescence count rates in single crystalline diamond in excess of 10${}^6$ counts per second. We show that our nano-fabrication procedure conserves the preferential alignment as well as the spin coherence of the NVs in our structures. Our results will enable a new generation of highly sensitive probes for NV magnetometry and pave the way toward photonic crystals with optimal orientation of the NV center's emission dipole.
\end{abstract}

\maketitle

%Motivation and background
Diamond nano-structures, including diamond nanopillars for single-photon wave-guiding,\cite{Babinec2010} diamond scanning probes for magnetometry,\cite{Maletinsky2012} diamond cantilevers for force-sensing,\cite{Ovartchaiyapong2012} or photonic crystals for the efficient cavity coupling of color centers \cite{Faraon2012,Riedrichmoeller2011} have been realized recently. A very promising, scalable and robust approach to engineer diamond-based quantum-devices is top-down nano-fabrication starting from single-crystalline diamond.\cite{Aharonovich2011b,Loncar2013} Without exception, the above mentioned structures have been fabricated on the $\left(100\right)$-facet of single crystalline diamonds -- the most common crystalline orientation and currently the only commercially available type of high-purity diamond material.
The functionality of these quantum devices mostly relies on the negatively charged nitrogen-vacancy defect complex in diamond (NV center). NV centers in diamond align along one of the four equivalent  $\left<111\right>$ crystal-directions and thus lie at an angle of $53^{\circ}$ with respect to $\left<100\right>$. This direction forms an important symmetry axis for most of the devices mentioned above\cite{Babinec2010, Maletinsky2012,Faraon2012,Riedrichmoeller2011} and the resulting oblique orientation of the NV represents a significant drawback in many cases. For instance, the photon collection efficiency for NV centers in nanopillars was predicted to be optimal if the emitting dipole is oriented perpendicularly to the pillar's axis;\cite{Hausmann2010} a situation which is only achieved when the pillar as well as the NV center are aligned along the same $\left<111\right>$ axis.  Similarly, for scanning NV magnetometry, $\left(111\right)$-oriented scanning probes, for which the NV axis can be perpendicular to the sample surface, provide significant improvements compared to existing devices, both in terms of magnetic field sensitivity and ease of interpretation of magnetometry data.\cite{Rondin2013} Using nano-cavities in photonic crystals fabricated from single crystalline diamond, the spontaneous emission of color centers can be significantly enhanced.\cite{Faraon2012,Riedrichmoeller2011,Hausmann2012} Here, the coupling of a color center to the cavity mode crucially depends on the alignment of the center's dipole moment to the cavity's electric field.\cite{Faraon2012} An orientation of this dipole in the plane of the photonic crystal is optimal and is, for the NV center, generally only achievable for $\left(111\right)$-oriented diamond.  Despite all these advantages, no $\left(111\right)$-oriented diamond nano-structures have been demonstrated so far. In particular, it is unclear if established diamond nano-fabrication approaches for such structures can be applied to (111)-oriented starting material. Very recently, high-quality, high-purity $\left(111\right)$-oriented chemical vapor deposition (CVD) diamond has become available.\cite{Tallaire2014} Single NV centers created during CVD growth in this material show very promising properties:\cite{Lesik2014,Michl2014} almost all NV centers align preferentially along the [111] growth axis and show long spin coherence times $T_2>100$ $\mu$s. Consequently, it is highly desirable to demonstrate that diamond nano-structures containing single NV centers can be fabricated in this material.

%Summary
We here demonstrate the fabrication of nano-photonic structures (nanopillars) from such $\left(111\right)$-oriented, single-crystalline CVD diamond. We characterize the photonic properties of native NV centers in these nanopillars and measure saturated fluorescence count rates
exceeding $10^6$ counts per second (cps), which provide evidence for a high collection efficiency for NV centers oriented along the nanopillar axis. Furthermore, we address the photonic properties of the pillars by carrying out finite differential time domain (FDTD) simulations. Importantly, our nano-fabrication procedure conserves the spin properties as well as the preferential alignment of the NV centers, which we demonstrate through electron spin resonance and coherent spin manipulation experiments.
\begin{figure}
\includegraphics[width=8.6cm]{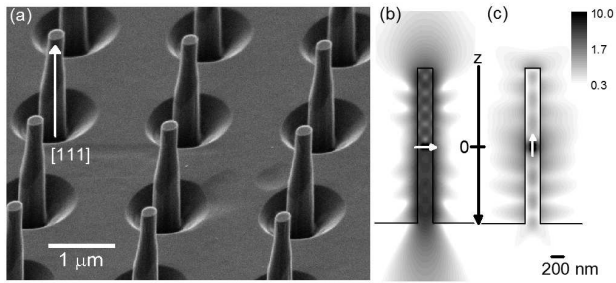}
\caption{(a) Scanning electron microscopy image of diamond nanopillars fabricated on a (111) oriented single-crystalline diamond sample. Note the smooth surfaces in-between the pillars. (b)+(c) FDTD simulations visualizing wave-guiding of the emission from a single dipole oriented perpendicularly to or along the nanopillar axis ($\lambda=$ 700 nm, images show magnitude of E field; white, solid arrows depict dipole orientation). The simulation has been performed for a cylindrical pillar with 230 nm diameter and 2.3 $\mu$m length. Note the enhanced coupling efficiency of a dipole oriented perpendicularly to the pillars axis. \label{FigPillarsSEM} }
\end{figure}

%Fabrication procedure and sample characterisation
Our fabrication recipe is based on previously published procedures for fabricating large, regular arrays of diamond nanopillars.\cite{Hausmann2010,Babinec2010} In particular, we are using e-beam lithography (30 keV) in order to pattern cylindrical etch-masks with approximately $200~$nm diameter and $550-600~$nm height into a layer of FOX-16 negative electron beam resist (Dow Corning).\cite{supplmat2011} A 2 nm titanium layer is deposited onto the diamond prior to resist spinning as an adhesion promoter. Before etching the diamond, the adhesion layer is removed using a short argon sputtering process. The developed mask is transferred into the diamond using inductively coupled plasma reactive ion etching (ICP-RIE, Sentech SI 500) to form the diamond nanopillars. The etch plasma is optimized to create vertical, smooth sidewalls of the pillars as well as a smooth surface in-between the pillars. To this end, we use a plasma containing 50\% argon and oxygen respectively (gas flow 50 sccm each). The pressure is set to 0.5 Pa, the ICP source is operated at a power of 500 W together with a bias power of 200 W. The observed etch rate on the $\left(111\right)$-oriented diamond was $260~$nm/min.
Our pillars were fabricated on the as-grown, cleaned surface of the $\left(111\right)$-oriented single-crystalline CVD diamond investigated in Ref.\ \onlinecite{Lesik2014}. As demonstrated there,\cite{Lesik2014} the NV centers in this sample exhibit near perfect orientation along the growth direction, i.e.\ out of the four possible orientations ($[111], [\overline{1}\,\overline{1} 1], [\overline{1}1\overline{1}], [1\overline{1}\,\overline{1}]$) for NV centers, they only occupy the subset along [111] due to the particular dynamics of diamond growth along $\left<111\right>$.

%Fabrication results
Figure \ref{FigPillarsSEM}(a) displays a scanning electron microscope image of typical diamond nanopillars which we fabricate on the $\left(111\right)$-oriented, single-crystalline diamond substrate.  Despite the previously reported high density of extended defects in this type of sample,\cite{Tallaire2014} the etched surface is smooth apart from a low density of triangular etch pits.\cite{supplmat2011} The etched pillars show an almost straight shape with a slight asymmetry. The length and diameter of the pillars are approximately 2.3 $\mu$m and 230 nm respectively. Additionally and in order to demonstrate the wider applicability of our nano-fabrication recipe, we also fabricate nanopillars on a polished, high purity, polycrystalline diamond sample. Here, we also obtain smooth surfaces with uniformly tall pillars on different grains of the polycrystalline material and only minor preferential etching of grain boundaries.\cite{supplmat2011}

\begin{figure}
\includegraphics[width = 8.6cm]{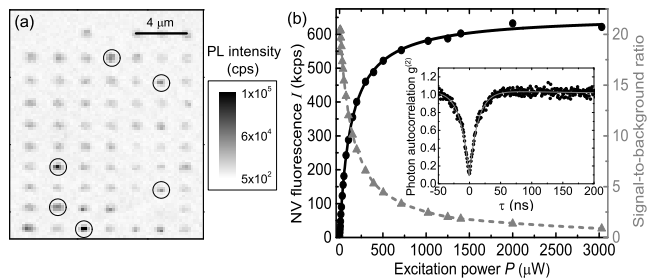}
\caption{ (a) Confocal image of an array of diamond nanopillars. Several pillars with single NV centers (g$^{(2)}$(0)=0.2-0.5) are marked with black, solid circles. The measurement was performed at 120 $\mu$W of excitation power.  (b) Signal-to-background ratio (SBR) and background corrected saturation curve for a single NV center. Inset: g$^{(2)}$-function of a single NV center. We find g$^{(2)}$(0)=0.12 without background correction, consistent with a probability of 96 \% that the detected photon stems from a single center or a SBR of approximately 20. The data was recorded at 20 $\mu$W excitation power.  \label{FigPhotonics}}
\end{figure}
%Confocal characterization
We characterize the diamond nanopillars using confocal microscopy (NA 0.8, excitation cw at 532 nm). Our setup is equipped with microwave control electronics to perform optically detected electron spin resonance experiments (ESR) on NV centers,\cite{Gruber1997}  a spectrometer (Acton SP2500, 300 grooves/mm grating) and correlation electronics (Fast ComTec, P7889) to perform second order autocorrelation measurements (g$^{(2)}$) of the fluorescence. For saturation measurements, an avalanche photodiode (Laser Components, Count-250C) with a quantum efficiency of 83\% at 670 nm is employed.

Figure \ref{FigPhotonics}(a) shows a confocal scan of an array of $[111]$-nanopillars. To identify single NV centers, we record the g$^{(2)}$-function of the fluorescence as displayed in the inset of Fig.\ \ref{FigPhotonics}(b) for a single NV center. Approximately 10-20\% of the pillars contain single NV centers as witnessed by a pronounced antibunching dip in the g$^{(2)}$-function with g$^{(2)}$(0)  significantly below 0.5. This probability to find a single NV center in a pillar is consistent with an NV density of 1.5-3 NVs/$\mathrm{\mu m^3}$ which is comparable to the approximate density of native NV centers of 1 NV/$\mathrm{\mu m^3}$ reported in Ref.\ \onlinecite{Lesik2014}. The fit (solid line) in the inset of Fig.\ \ref{FigPhotonics}(b) uses the g$^{(2)}$-function of a three level system, including uncorrelated background.\cite{Brouri2000} It is consistent with a signal-to-background ratio of approximately 20 and thus confirms very pure single photon emission from our devices. 
Figure \ref{FigPhotonics}(a) also shows that even the pillars that do not contain NV centers exhibit a certain amount of fluorescence which we assign to broadband background (spectral range 650-800 nm) stemming from the starting material.\cite{Tallaire2014} We note that this weak background luminescence does not significantly affect the purity of the NV center emission from the nanopillars which displays the typical fluorescence spectrum of single NV centers including a pronounced zero-phonon-line.\cite{supplmat2011} 

\begin{figure}
\includegraphics[width =8.6cm]{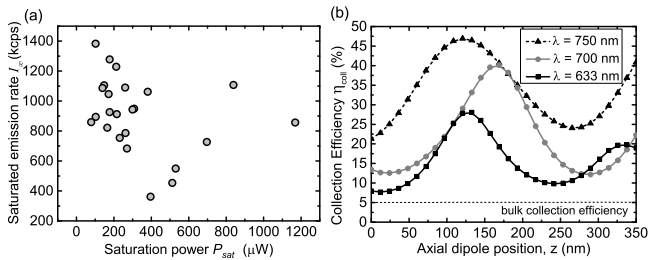}
\caption{(a) Saturation power $P_{\mathrm{sat}}$ and maximum saturated emission rates $I_\infty$ obtained for 25 single NV centers. (b) Collection efficiency $\eta_{\mathrm{coll}}$ extracted from FDTD simulations for an emitting dipole perpendicular to the nanopillar axis. $\eta_{\mathrm{coll}}$ is shown for three different wavelengths. The emitting dipole is moved along the pillar axis (coordinate z, as defined in Fig.\ \ref{FigPillarsSEM}) 350 nm toward the bulk diamond substrate starting from the middle of the pillar.   \label{Figsim}}
\end{figure}

%Count rate and saturation curves
To characterize the photonic properties of our nanopillar devices, we perform saturation-measurements on a subset of  pillars which contain single [111]-oriented NV centers. To separate background emission and fluorescence from single NV centers, we fit the measured fluorescence intensity $I$ in dependence of the excitation power $P$ using \begin{equation}
I(P)=I_\infty\frac{P}{P+P_{\mathrm{sat}}}+c_{b}P.
\end{equation}
$P_{\mathrm{sat}}$ is the saturation power, $I_\infty$ the fully saturated emission rate of the NV center and the last term $c_{b}P$ accounts for background.\cite{Kurtsiefer2000} A characteristic background corrected saturation curve for a single NV is displayed in Fig.\ \ref{FigPhotonics}(b).  Figure \ref{Figsim}(a) summarizes $I_\infty$ and $P_{\mathrm{sat}}$ obtained for 25 single NV centers. We find an average $I_\infty$ of 910$\pm$250 kcps and several single NVs for which $I_\infty$ significantly exceeds \mbox{1 Mcps}. The saturation power $P_{\mathrm{sat}}$ is below 400 $\mu$W for the majority of the pillars. Using the saturation curves, we can also estimate the signal-to-background ratio (SBR) for the fluorescence of single NV centers which varies significantly between the pillars and can be as high as 60 for low excitation powers.\cite{supplmat2011} For an excitation power of 100 $\mu$W, the majority of pillars investigated has an SBR of better than 15. Consequently, compared to the investigation in bulk, the SBR is enhanced by a factor of 5-10.\cite{Lesik2014} Thus, the nanopillars enable very pure single photon emission from NV centers, even in a diamond material exhibiting broad background fluorescence.

In our data in Fig.\ \ref{Figsim}(a), we observe a significant spread of $P_{\mathrm{sat}}$ and $I_\infty$, which we assign to different locations of the individual NVs within the nanopillars. Indeed, the NV centers are randomly distributed (radially and axially) in the pillars due to their formation from residual nitrogen during the diamond growth. Excitation of the NV center as well as collection of the fluorescence are both mediated by the pillar's photonic modes [see Fig.\ \ref{FigPillarsSEM}(b)+(c)] and are thus expected to vary depending on the spatial position of the NV center in the pillar.\cite{Babinec2010} To investigate the influence of the spatial position, we simulate the electromagnetic fields in the pillar using the FDTD method implemented using commercial software (Lumerical FDTD solutions). First, we simulate the Gaussian excitation field in the nanopillar (cylindrical pillar, diameter $d=$230 nm, 2.3 $\mu$m length, refractive index 2.418). From this, we obtain that NV centers close to the pillar's axis experience an excitation field enhanced by 60\% compared to centers close to the pillar's side walls,\cite{supplmat2011} thus directly accounting for a spread in the observed $P_{\mathrm{sat}}$.  To estimate the expected spread of $I_\infty$, we use a single radiating dipole perpendicular to the pillar axis for the simulation [Fig.\ \ref{FigPillarsSEM}(b)]. We choose a broadband dipole source with a bandwidth roughly matching the NV center's emission spectrum. From these simulations, we obtain the far field of the pillars. The source is placed in 20 different positions along the nanopillar and we extract the collection efficiency $\eta_{\mathrm{coll}}$ (numerical aperture 0.8) for 10 different wavelengths $\lambda$ within the NV emission spectrum for each position. As observed in Fig.\ \ref{Figsim}(b), the axial position of the NV strongly influences $\eta_{\mathrm{coll}}$, thus accounting for the observed spread of $I_\infty$. The maximum $\eta_{\mathrm{coll}}$ exceeds the bulk value ($\approx 5\%$) by almost an order of magnitude. We note that according to our simulations and previous discussions in Ref.\ \onlinecite{Friedler2009}, the radial position does not significantly affect $\eta_{\mathrm{coll}}$. For the lifetime of the emitting dipole however, a placement off axis, in contrast to on axis placement, leads to a suppression of spontaneous emission (maximum simulated suppression: emission rate $\approx 30\%$ bulk value) due to changes in the local density of states. Taking into account possible non-radiative decays for the NV center, however, it remains questionable if lifetime changes can be observed. As the NV center's emission spans a wavelength range of around 100 nm, it is also instructive to investigate the wavelength dependence of $\eta_{\mathrm{coll}}$. Fig.\ \ref{Figsim}(b) shows a trend for enhanced $\eta_{\mathrm{coll}}$ for longer wavelengths $\lambda$. This is consistent with earlier simulations,\cite{Friedler2009} which demonstrated higher transmission of the guided mode through the pillar's top facet for smaller effective diameters ($d$/$\lambda$). For comparison and to illustrate the advantages of a [111] oriented NV center, we simulate a dipole parallel to the pillar axis [see Fig.\ \ref{FigPillarsSEM}(c)]. From our simulations, we conclude that for our collection NA, an NV dipole oriented orthogonally to the wire axis only gives a slight improvement  ($\approx30-40\%$, depending on position in the wire) in collected fluorescence compared to a dipole parallel to the wire. The high NV fluorescence count rates we observe are thus caused by a combination of this enhancement and improvements of our setup's efficiency compared to previous approaches.

\begin{figure}
\includegraphics[width = 8.6cm]{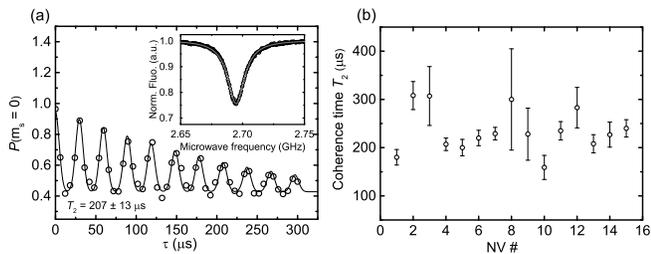}
\caption{ (a) Spin echo measurement performed on a [111]-oriented NV center in a nanopillar. The solid line is a fit with a sum of Gaussian peaks modulated by a decay envelope.\cite{Childress2006a} The inset shows the ESR resonance of a single NV center, displaying a contrast of 25\% for the ESR measurement under continuous laser excitation. (b) Summary of spin coherence times $T_2$ for 15 NV centers in nanopillars.  \label{FigSpin}}
\end{figure}

%Orientation of NVs and discussion of lost preferential alignment
The starting material for the fabrication of the nanopillars has been found to contain NV centers which orient along the $[111]$ growth axis of the sample with a probability of 97\%.\cite{Lesik2014} To verify that the nano-fabrication procedure does not influence the preferential alignment, we apply a static magnetic field aligned with $[111]$-oriented NV centers. In this configuration, ESR spectra reveal the orientation of the NV centers. From the ESR measurements, we conclude that all 25 single NV centers we investigated are oriented along $[111]$ and the preferential alignment is maintained in the nanopillars. All investigated centers exhibit a very pronounced optically detected ESR resonance, whose contrast under continuous laser excitation reaches 25\%.
%Coherence of NVs
Finally, we verify that our fabrication procedure does not affect spin coherence properties of NV centers in our sample. To that end, we perform Hahn-Echo measurements on a subset of the nanopillars containing single NV centers; the results are summarized in Fig.\ \ref{FigSpin}. Before nano-fabrication of the pillars, a $T_2$ time of 190 $\mu$s has been measured for NV centers in the bulk diamond.\cite{Lesik2014} For single NVs in the nanopillars, we find $T_2$ times between 160 and 310 $\mu$s with an average $T_2$ time of 235$\pm$46 $\mu$s. 

In summary, we fabricated for the first time nano-photonic structures, namely nanopillars, with approximately 2 $\mu$m length and 200 nm diameter, on high-purity, $\left(111\right)$-oriented single-crystalline CVD diamond. We characterize single native NV centers in these nanopillars and find high saturated fluorescence count rates around 10${}^6$ cps. The nano-fabrication procedure conserves the preferential alignment of the centers as well as the spin coherence properties. This demonstration of nano-device fabrication on $\left(111\right)$-oriented diamond paves the way toward diamond scanning probes for magnetometry and photonic crystals with optimal orientation of the NV center's emission dipole. Ultimately, isotopically pure diamond growth in combination with the formation of NV centers in a nanometer thin layer ($\delta$-doping) can lead to devices with a controlled depth of the NV centers as well as ultra-long spin coherence times.\cite{Balasubramanian2009,Ohno2012}  

%Acknowledgements
We gratefully acknowledge financial support through the NCCR QSIT, a competence center funded by the Swiss NSF, and through SNF Grant No. 200021\_143697/1. This research has been partially funded by the European Commission's 7.\ Framework Program (FP7/2007-2013) under grant agreement number 611143 (DIADEMS).

% Create the reference section using BibTeX: Bibliography appended in the end
%

\clearpage
\textbf{Supplementary material for manuscript Photonic nano-structures on $\left(111\right)$-oriented diamond}

\textit{This supplementary material summarizes additional technical details on chemical vapor deposition growth of the employed diamond sample and the nanofabrication process. It furthermore presents additional data on spectral and photonic properties of the NV center in the $\left(111\right)$-oriented nanopillars as well a details on the FDTD simulations.}

\textbf{Details on chemical vapor deposition (CVD) growth of the (111) oriented diamond sample:}

Growth was carried out on (111) plates cleaved and polished with a slight off-angle (2-5$^\circ$) from High Pressure High Temperature (HPHT) synthetic diamond crystals. The crystals with a high crystalline quality and relatively low impurity amount were grown by the temperature gradient method using a split-sphere (BARS) apparatus at the Institute of Geology and Mineralogy of Novosibirsk. After acid cleaning, they were introduced in a home-made resonant-cavity reactor chamber able to operate at fairly high pressures and microwave powers (typically 250 mbar, 3500 W). A high purity mixture of H$_2$/CH$_4$ was used (98/2) at a temperature of about 1050$^\circ$C. The background N$_2$ concentration is believed to be under 1 ppm in the gas phase under these conditions. The use of both, low methane concentration and high temperature, was found to strongly improve diamond crystal morphologies by efficiently suppressing twinning that easily occurs on this orientation. In this way, it was possible to grow films with thicknesses well over 50 $\mathrm{\mu}m$ as the one used in this study.

\textbf{Further details on nanofabrication:}

\begin{figure}
\includegraphics[width=7cm]{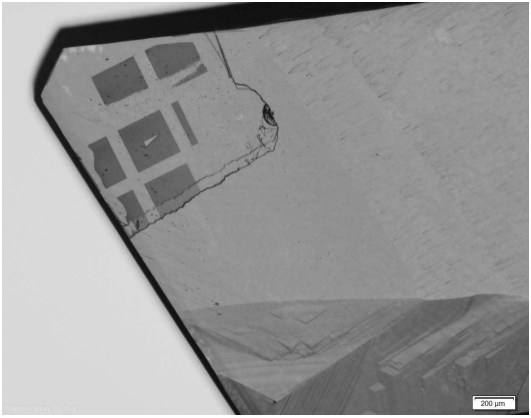}
\caption{Light microscope image of the $\left(111\right)$-oriented CVD diamond sample. The structured region has been placed in the very smooth area of the sample. }
\label{FigSamplepic}
\end{figure}

As mentioned in the manuscript, we use the sample characterized in Ref.\ \onlinecite{Lesik2014}. Figure \ref{FigSamplepic} shows a light microscope image of the sample. Great care has been taken to place the processed area fully in the smooth area on the sample close to the pre-characterized region.
\begin{figure}
\includegraphics[width=8.6cm]{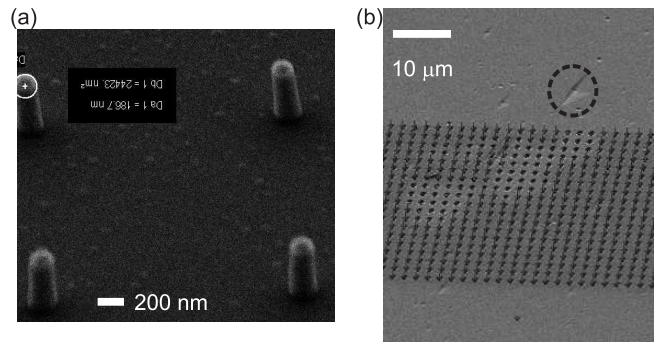}
\caption{(a) FOX mask pillars, the surface in-between the mask reveals a slight roughness that can be attributed to the titanium adhesion layer. (b) Array of $\left(111\right)$-oriented nanopillars, around the pattern a low density of triangular etch pits is visible (one etch pit has been highlighted with a dashed circle). \label{Figmask}} 
\end{figure}
We are using e-beam lithography (30 keV) in order to pattern cylindrical etch-masks with approx. $200~$nm diameter and $550-600~$nm height into a layer of FOX-16 negative electron beam resist (Dow Corning).\cite{supplmat2011} A 2 nm titanium layer is deposited onto the diamond using electron beam evaporation prior to resist spinning as an adhesion promoter. The resist is spun onto the diamond with 6000 rpm. We use two layers of FOX. Spin-coating of the first layer is followed by a 5 mins bake at 90$^\circ$C. After spin-coating the second layer, 10 mins baking at 90$^\circ$C is employed. E-beam lithography has been performed using a Zeiss 45 Supra Scanning electron microscope (SEM) equipped with an Elphy Raith lithography extension. The acceleration voltage for the lithography is 30 keV, the dose for the pillars was 1700 $\mu$C/cm$^2$. Following e-beam lithography, the mask is developed in 25\% TMAH solution for 20 s and rinsed in deionized water prior to etching. Figure \ref{Figmask}(a) shows an SEM image of the FOX mask pillars before etching.  Prior to transferring the mask into the diamond using the Ar/O$_2$ plasma described in the manuscript, we employ a 45 s Ar sputtering process in the same etch reactor to remove the titanium adhesion layer that oxidizes in-between the e-beam lithography and the etching steps (plasma parameters: 0.4 Pa, 50 sccm Ar, 500 W ICP, 300 W bias power). After finishing the etching process, the FOX mask as well as the residual titanium adhesion layer are removed using an overnight treatment in Buffered Oxide etch solution [hydrofluoric acid (HF) and ammonium fluoride (NH$_4$F), approx. 1:20]. Subsequently, the sample is cleaned in a boiling triacid clean (Sulphuric acid, Nitric acid, perchloric acid, 1:1:1) to yield a defined oxygen terminated low fluorescence surface. Finally, the sample is rinsed in de-ionized water and solvents (acetone, ethanol, isopropanol) and blow dried with nitrogen.
\begin{figure}
\includegraphics[width=8.6cm]{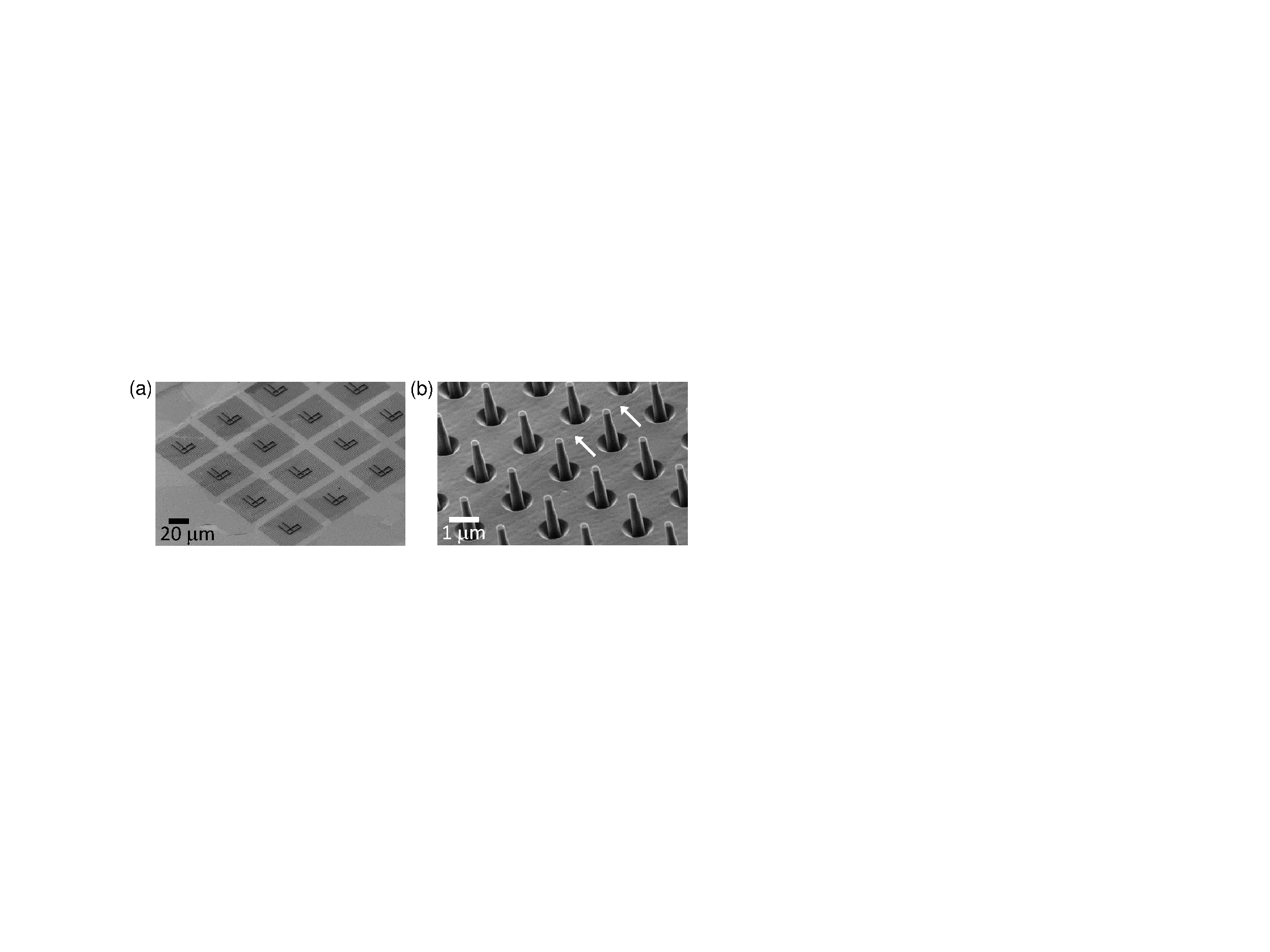}
\caption{(a) Large array of nanopillars formed on the polished surface of a high purity polycrystalline diamond film. The grains in the diamond film are visible as regions with different contrast in the SEM. (b) Detailed image of nanopillars on polycrystalline film. White arrows mark a position where a grain boundary is slightly visible. Note that despite the etch depth of almost 2 $\mu$m, only a very small etch pit is found at the position of the grain boundary. \label{PCDpillars}}
\end{figure}

The surface between the etched wires is mostly smooth; only a very low density of triangular etch pits is formed. The etch pits are visible as dark spots in Fig.\ \ref{FigSamplepic} and in more detail in the electron microscope image in Fig.\ \ref{Figmask}(b). To demonstrate the wider applicability of our nanofabrication recipe, we also fabricate nanopillars on a polished high purity, polycrystalline diamond sample (Element Six, electronic grade CVD diamond, see Fig.\ \ref{PCDpillars}).  Reference \onlinecite{Markham2010} finds that the properties of NV centers in polycrystalline diamond material can almost match those in single crystalline material. Moreover, polycrystalline material can be fabricated on larger areas and might thus be an alternative to single crystalline material given the fabrication strategy used is suitable for polycrystalline material and the crystal orientation in the region of interest can be measured. The imaged area on the polycrystalline sample comprises several crystallites, as indicated by the varying SEM image contrast from the different regions. Our fabrication process yields a smooth surface in the complete etched area as well as uniformly high pillars on different crystallites. Polycrystalline diamond films often exhibit a preferential orientation,\cite{Balmer2009} however, Electron Backscatter Diffraction (EBSD) measurements on the polycrystalline diamond material we used reveal no preferential alignment. Thus, the etch results on the polycrystalline material indicate that grains with significantly different orientation etch almost uniformly. Furthermore, it is very promising that the grain boundaries in the polycrystalline material are almost not attacked during the etching and no etch pits are formed [see Fig.\ \ref{PCDpillars}(b)].

\textbf{Further data on nanopillar characterization: Photonic properties}

\begin{figure}
\includegraphics[width=6.5cm]{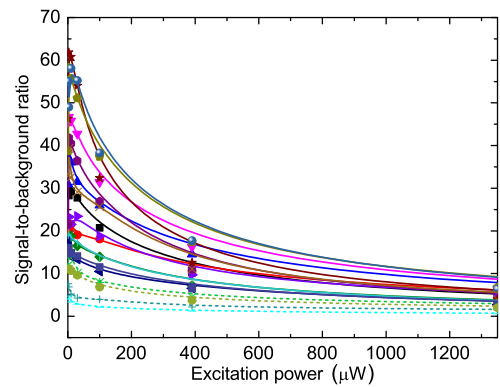}
\caption{Signal-to-background ratio for 18 individual NV centers in nanopillars. The dashed lines correspond to nanopillars with very high saturation power P$_{sat}$ or very high background emission. The background emission varies between 116 and 360 cps/$\mu$W.}
\label{FigSignaltonoise}
\end{figure}

As described in the manuscript, the nanopillars allow for an observation of single NV centers with very high signal-to-background ratio (SBR). Figure \ref{FigSignaltonoise} summarizes the SBR obtained for 18 single NV centers from saturation measurements. At low excitation power, the SBR can be as high as 60. Naturally, above saturation of the single emitter it drops significantly. Single NV centers with comparably low SBR (dashed lines in Fig.\ \ref{FigSignaltonoise}) have been found to exhibit the highest saturation powers P$_{sat}$ observed and thus indicate a less efficient excitation of the NV centers and consequently a high excitation of the background fluorescence.

\textbf{Further data on nanopillar characterization: Spectra of single NV centers}
\begin{figure}
\includegraphics[width=6.5cm]{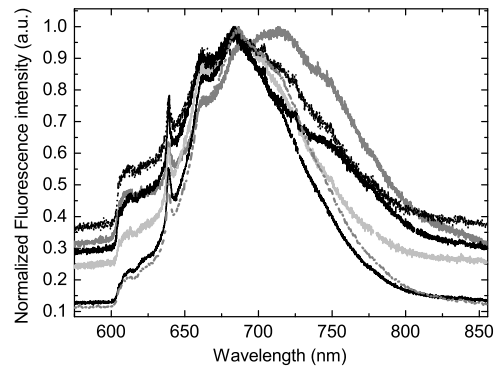}
\caption{Spectra of single NV centers (g${^(2)}$(0)=0.2-0.5). The ZPL is clearly visible in all spectra, a variation of the spectrum throughout the sidebands can be attributed either to variations in linear electron phonon coupling between different centers or to a wavelength-dependent coupling efficiency to the nanopillar.}
\label{FigSpectra}
\end{figure}
Fig.\ \ref{FigSpectra} summarizes spectra of single NV centers. We find that the purely electronic transition (zero-phonon-line, ZPL) is clearly visible in all spectra. Its center wavelength is found between 638.7 and 639 nm; the width varies between 2.2 and 2.6 nm. Despite the very uniform appearance of the ZPL, the sideband spectrum of the NV centers varies between individual centers. The variation in the sideband spectra might arise either from variations in linear electron-phonon-coupling between different centers\cite{Tizei2012} or from a wavelength-dependent coupling-efficiency to the nanopillar which leads to distortions in the recorded sideband-spectrum.

\textbf{Further information on FDTD simulations: Mode pattern of the excitation laser }

\begin{figure}
\includegraphics[width=8cm]{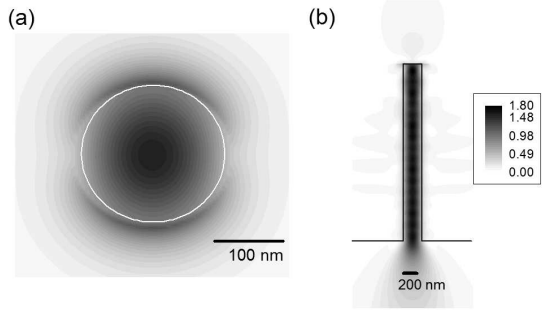}
\caption{Excitation of the diamond nanopillars (cylindrical pillar, diameter $d=$230 nm, 2.3 $\mu$m length) using 532 nm light focused to a Gaussian spot with 400 nm FWHM. Note that the color bar is valid for part (a) and (b) of the figure. (a) Cut through the wire in the middle of the wire. (b) Cut through the wire's mode profile. }
\label{Figexc}
\end{figure}

Figure \ref{Figexc} shows the coupling of laser light with a wavelength of 532 nm from a Gaussian source into the diamond nanopillar. The light is well guided in the nanopillar, ideal coupling into the pillars is achieved when the laser is focused onto the top facet of the pillar. NV centers close to the wire's axis experience an excitation field enhanced by 60\% compared to centers close to the wire's side walls, thus straightforwardly accounting for a spread in  $P_{\mathrm{sat}}$. 
We performed the optimized calculations with an 8-nm grid, running the simulation until the power in the simulation region falls below $10^{-5}$ of the injected power. Simulation region was wide enough to account for the large NA used in the experiment.

\end{document}